# Material-based Non-neural Analogues of Lateral Inhibition: A Multi-agent Approach

Jeff Dale Jones

Centre for Unconventional Computing
University of the West of England
Coldharbour Lane
Bristol, BS16 1QY, UK.
`jeff.jones@uwe.ac.uk`

**Abstract.** Lateral Inhibition (LI) phenomena occur in a wide range of sensory modalities and are most famously described in the human visual system. In LI the activity of a stimulated neuron is itself excited and suppresses the activity of its local neighbours via inhibitory connections, increasing the contrast between spatial environmental stimuli. Simple organisms, such as the single-celled slime mould *Physarum polycephalum* possess no neural tissue yet, despite this, are known to exhibit complex computational behaviour. Could simple organisms such as slime mould approximate LI without recourse to neural tissue? We describe a model whereby LI can emerge without explicit inhibitory wiring, using only bulk transport effects. We use a multi-agent virtual material model of slime mould to reproduce the characteristic contrast amplification response of LI using excitation via attractant stimuli. Restoration of baseline activity occurs when the stimuli are removed. We also explore an opposite counterpart behaviour, Lateral Activation (LA), using repellent stimuli. These preliminary results suggest that simple organisms without neural tissue may approximate sensory contrast enhancement using alternative analogues of LI and suggests novel approaches towards generating collective contrast enhancement in distributed computing and robotic devices.

Living organisms perceive their environment with a wide variety of special senses. Enhancing the contrast in the stream of information from these senses allows organisms to discriminate between small changes in signal level, potentially enhancing survivability. Lateral Inhibition (LI) is a neural mechanism which enhances the activity of neurons directly exposed to excitatory stimuli whilst suppressing the activity of their near neighbours. LI phenomena have been described in auditory, somatosensory and olfactory senses, but are most famously described in the visual systems of a wide range of organisms [2, 5, 4].

LI mechanisms are an effective and efficient means of enhancing environmental perception in organisms ranging from the most primitive to the most complex. But can such mechanisms exist in organisms which do not possess any nervous system? For example, the giant amoeboid true Slime Mould *Physarum polycephalum* is a single-celled organism possessing no neural tissue but which (in its vegetative plasmodium stage) exhibits a complex range of biological and computational behaviours (for a comprehensive overview of its abilities see [1]). The plasmodium of *P. polycephalum* is comprised of an adaptive complex gel/sol transport network. The ectoplasmic gel phase is composed of a sponge-like matrix of actin and myosin fibres through which the endoplasmic sol flows, transported by spontaneous and self-organised oscillatory contractions. The organism behaves as a distributed computing *material*, capable of responding to a wide range of spatially represented stimuli. How can slime mould perform such complex feats with such simple components? In this abstract we describe a collective mechanism by which sensory contrast enhancement phenomena analogous to LI phenomena can emerge in unorganised non-neural systems. We use a multi-agent particle based model of slime mould which replicates the self-organised network formation and adaptation of slime mould (see [3] for a detailed description). We examine the emergence of LI in response to attractant stimuli and the opposite phenomenon, Lateral Activation, is described in response to adverse stimuli (simulated light irradiation).

We initialised the model virtual plasmodium comprising 8000 particles within a 300 × 100 pixel tube-like horizontal arena bordered by inhabitable areas on the top and bottom and open ended left and right edges. Periodic boundary conditions were enforced. We measured population density across the arena by counting the number of particles in the Y-axis for each X-axis position. We recorded population density every 10 scheduler steps. After initialisation the population, constrained by the architecture of the arena, formed a single tube with relatively uniform population density (Fig. 1,a).

An attractant stimulus was presented to the virtual plasmodium after 1000 scheduler steps by projecting chemoattractant into the middle-third habitable section of the arena (Fig. 1,b). The attractant stimulus caused increased flux of particles into the stimulus area, an increase in population density in this area, and a corresponding decrease in density outside the stimulus region (Fig. 1,c). Upon removal of the stimulus after 4000 steps the population was no longer preferentially attracted to the central region and the tube adapted its shape in response to the uniform chemoattractant profile (Fig. 1,d). The population density eventually returned to uniform density across the arena (Fig. 1,e). The density cross-section profiles (Fig. 1f-j) demonstrates the excitation effect within the attractant region whilst the neighbouring regions show an decrease in population density.

To examine the collective response to Adverse Stimuli (simulated exposure to illumination, which the *Physarum* plasmodium avoids), we used the same arena and illumination pattern but the attractant region was replaced with a region of simulated illumination. This is represented in the model by reducing the sensitivity of the sensor values in illuminated areas and reducing chemoattractant values in the exposed areas. Particles at the border of exposed areas preferentially moved to unexposed regions and the local coupling of particles resulted in collective flux away from the stimulus area (Fig. 2a-e). The density cross-section profiles (Fig. 2f-j) demonstrates the inhibition effect within the illuminated region whilst the unexposed neighbouring regions show an increase in population density. When the adverse stimulus was removed from the central region the population density re-normalised to a uniform level within 15000 steps. A space-time plot of the population density indicates how the changes in population density for both conditions are initiated at the stimulus boundaries (Fig. 3). Regions of increased density inside the attractant stimulus correspond to excitation and regions outside correspond to inhibition. For the adverse stimuli regions inside correspond to inhibition and regions outside correspond to excitation.

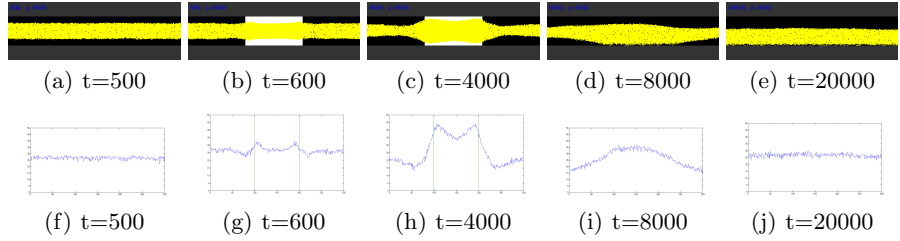

(a) t=500　　(b) t=600　　(c) t=4000　　(d) t=8000　　(e) t=20000

(f) t=500　　(g) t=600　　(h) t=4000　　(i) t=8000　　(j) t=20000

**Fig. 1.** Response of virtual plasmodium flux to attractant stimulus. a) population initialised within horizontal arena forms single tube, b) presentation of attractant stimulus bar (light area) results in flux towards stimulus area, c) population density is increased at stimulus region and reduced at unstimulated region, d) removal of stimulus results in adaptation to uniform attractant profile, e) uniform density is restored, f,g,h,i,j) cross-section plots of population density across the arena.

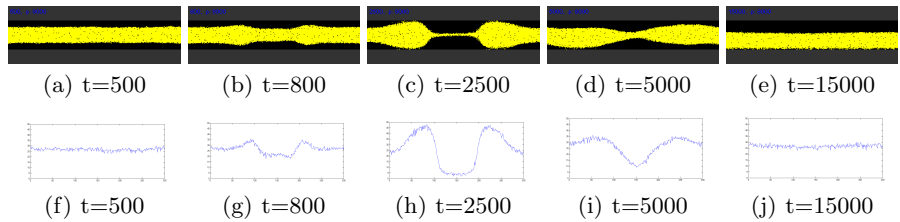

(a) t=500　　(b) t=800　　(c) t=2500　　(d) t=5000　　(e) t=15000

(f) t=500　　(g) t=800　　(h) t=2500　　(i) t=5000　　(j) t=15000

**Fig. 2.** Response of virtual plasmodium flux to simulated light irradiation. a) population initialised within horizontal arena forms single tube, b) presentation of simulated light irradiation (centre, not shown) results in flux away from irradiated area, c) population density is decreased at irradiated region and increased at unexposed region, d) removal of adverse stimulus results in increased flux to inner region, e) uniform density is restored, f,g,h,i,j) cross-section plots of population density across the arena.

We have demonstrated the results of scoping experiments into the generation of spatial contrast enhancement analogous to Lateral Inhibition in non-neural systems using a multi-agent model of slime mould *Physarum polycephalum*. The results show the classic LI contrast enhancement in response to attractant stimuli and its opposite counterpart behaviour (Lateral Activation) in response to adverse stimuli (simulated light irradiation). These effects are generated by bulk transport of the particles comprising the virtual material at the borders of stimuli projection, and restoration of uniform baseline activity (population density distribution) is established when stimuli are removed. The most notable feature of this approach is that LI phenomena can be approximated without explicit fixed inhibitory connections. This suggests possible mechanisms by which organisms without neural tissue may achieve sensory contrast enhancement. In the context of adaptive materials and robotics applications the mechanism illustrates how complex sensory behaviour can be distributed within an unorganised material (or robotic collective) itself. This allows greater freedom from having to pre-specify connectivity to implement sensory contrast enhancement and allows redundancy for individual faulty components. We hope that ongoing research may lead to other unorganised material approximations of complex neural functions such as brightness perception, spatial feature detectors, and direction discrimination.

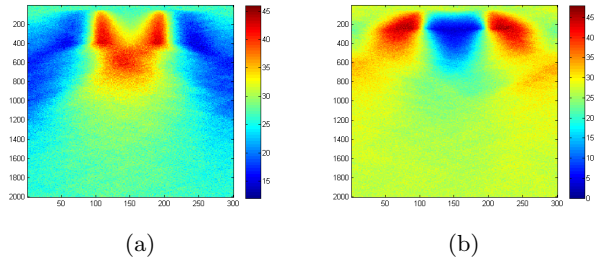

(a)               (b)

**Fig. 3.** Space-time plots of changing population density flux, time progresses downwards. Note that changes in density are initiated at borders of stimulus boundaries and the restoration of baseline activity after stimulus removal. a) Attractant condition reproducing LI response, b) Simulated light irradiation reproducing LA response.

This work was supported by the EU research project "Physarum Chip: Growing Computers from Slime Mould" (FP7 ICT Ref 316366)